\documentclass[ aip, amsmath,amssymb, reprint ]{revtex4-1}
\usepackage{graphicx}
\usepackage{dcolumn}
\usepackage{bm}
\usepackage[utf8]{inputenc}
\usepackage[T1]{fontenc}
\usepackage{mathptmx}
\usepackage{etoolbox}

\makeatletter
\def\@email#1#2{%
 \endgroup
 \patchcmd{\titleblock@produce}
  {\frontmatter@RRAPformat}
  {\frontmatter@RRAPformat{\produce@RRAP{*#1\href{mailto:#2}{#2}}}\frontmatter@RRAPformat}
  {}{}
}%
\makeatother
\begin{document}

\title{High-density liquid (HDL)  adsorption at the supercooled water/vapor interface and its possible relation to the second surface tension inflection point}
\author{Alexander Gorfer}
\affiliation{Faculty of Physics, University of Vienna, Boltzmanngasse 5, Wien A-1090, Austria}
\author{Christoph Dellago}
\affiliation{Faculty of Physics, University of Vienna, Boltzmanngasse 5, Wien A-1090, Austria}
\affiliation{Erwin Schrödinger Institute for Mathematics and Physics, Boltzmanngasse 9, 1090, Vienna,
Austria}
\author{Marcello Sega}
\affiliation{Department of Chemical Engineering, University College London, London WC1E 7JE, United Kingdom}
\email{m.sega@ucl.ac.uk}

\date{\today}

\begin{abstract}\noindent{}We investigate the properties of water along the liquid/vapor coexistence line in the supercooled regime down to the no-man's land. Extensive molecular dynamics simulations of the TIP4P/2005 liquid/vapor interface in the range 198 -- 348 K allow us to locate the second surface tension inflection point with high accuracy at 283$\pm$5 K, close to the temperature of maximum density (TMD). This temperature also coincides with the appearance of a density anomaly at the interface known as the apophysis. We relate the emergence of the apophysis to the observation of HDL water adsorption in the proximity of the liquid/vapor interface.

\end{abstract}
\maketitle

\section{Introduction}
The rich set of anomalous properties of water has kept it an active area of research, in particular regarding the thermodynamic and kinetic properties of the supercooled liquid state.  These properties are governing the freezing of water\cite{nistor_interface-limited_2014,fitzner_ice_2019} and are key in diverse fields like climatology~\cite{hu_occurrence_2010,murray_ice_2012} or cryobiology~\cite{mazur_cryobiology_1970,wilson_ice_2003}, but experiments  aimed at investigating the properties in the deeply supercooled region face an almost insurmountable challenge. Below the temperature of homogenous nucleation \(T_H\), frequently reported to be around 235 K~\cite{Manka2012}, water freezes faster than the time it takes to measure response functions. Therefore, most measurements become impossible. The region of temperatures below \(T_H\) has been named the no-man's land, and the properties of water in this region have become topics of intense debate~\cite{Palmer2018}.

One of the subjects of debate is the source of the drastic evolution of water's response functions in the supercooled regime. A popular explanation involves the presence of a liquid-liquid first order phase transition~\cite{Gallo2016} between high-density (HDL) and ice-like low-density liquid (LDL) water upon lowering the temperature. In this scenario, the associated critical point explains the divergent behavior of water's response functions inside the no-man's land. 

A second debated topic is the presence of a second, low-temperature inflection point of the surface tension as a function of temperature. Above the melting point, the surface tension of water follows with high accuracy the empirical equation of the International Association for the Properties of Water and Steam (IAPWS)~\cite{IAPWS2014,Vargaftik1983} 
\begin{equation}\label{Eq:IAPWS}
    \gamma = B \left(\frac{T_\text{c} - T}{T_\text{c}}\right)^\mu \left(1 + b \frac{T_\text{c} - T}{T_\text{c}}\right),
\end{equation}
where \(T_\text{c}\) = 647.096 K is the critical temperature and the remaining parameters are \(B = 235.8\) mN/m, \(b = -0.625\) and \(\mu = 1.256\). The IAPWS curve agrees with measured data well below the melting point~\cite{Hruby2014}. However, whether this agreement can continue deeper in the supercooled regime is controversial, as measurements of surface tension become very involved~\cite{Vins2015}. A recent experimental study~\cite{Vins2020} measured the surface tension of water down to about 241 K using the capillary rise method and reported a deviation from the IAPWS equation that opens the possibility of the existence of a second inflection point. Computational studies support the appearance of the second inflection point for several water models, including SPC/E~\cite{Lu2006, Wang2019}, TIP4P/2005~\cite{Wang2019} and WAIL~\cite{Rogers2016}.

\begin{figure}
    \begin{center}
    \includegraphics[width=.8\columnwidth]{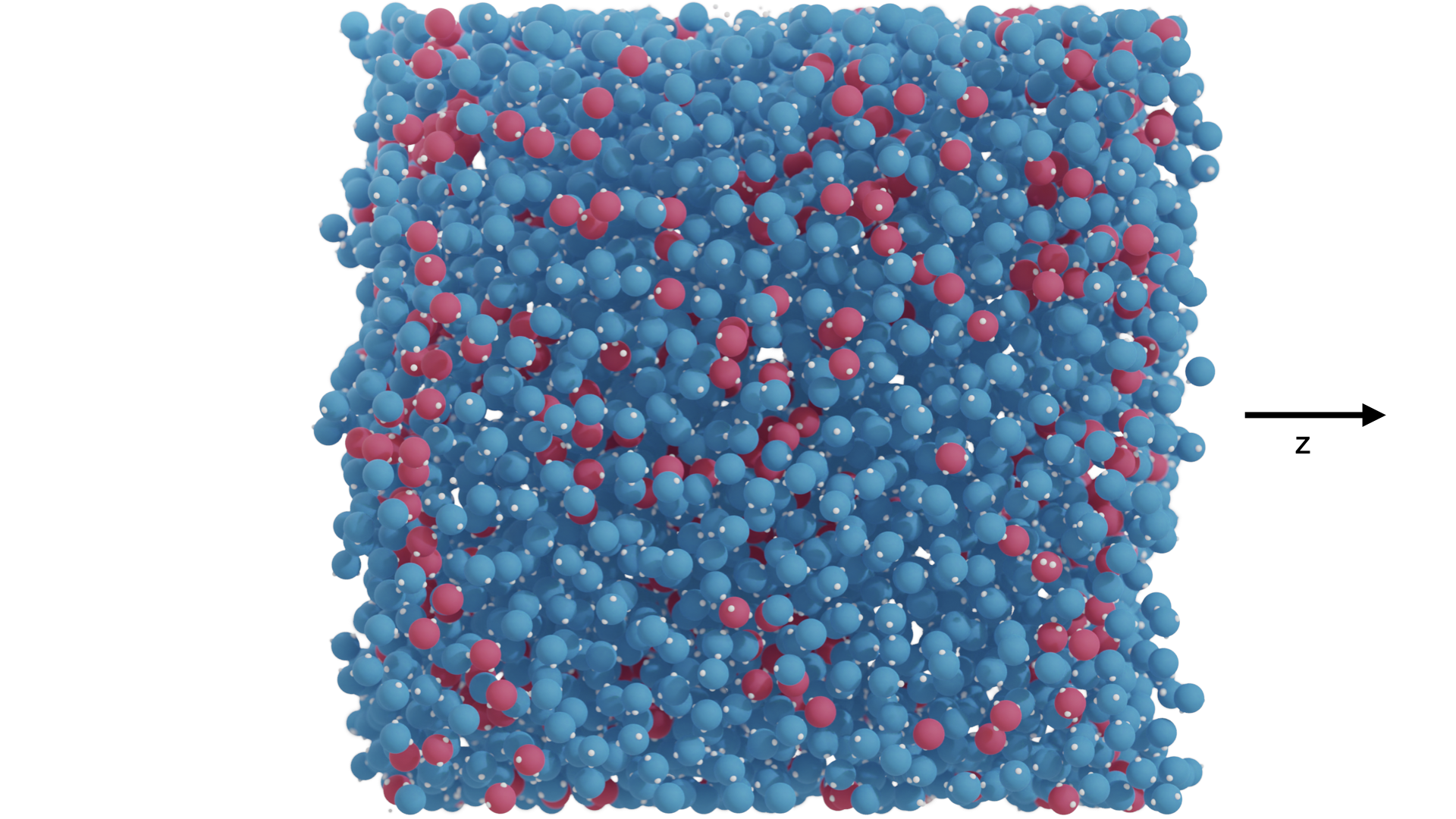}
    \end{center}
    \caption{Simulation snapshot taken at 198.15 K, showing HDL (red oxygen atoms) and LDL (blue oxygen atoms) molecules, where the accumulation of HDL molecules close to the liquid/vapor interface  is evident by simple visual inspection. The surface normal points along \(z\).}   \label{Fig:HDL_render}
\end{figure}

Simulations of explicit liquid/vapor coexistence of supercooled water have also exhibited a compact layer at the water/vapor interface that manifests itself as a shoulder in the density profiles \cite{Matsubara2007, Abe2014, Haji-Akbari2014, Haji-Akbari2017, Malek2018, Wang2019}. Wang and coworkers~\cite{Wang2019} suggested a connection of this feature to the second inflection point of the surface tension. Still, a systematic study of this anomaly and its links to other anomalous features has not been attempted so far. The outermost layer of the water/vapor interface is known to have a significantly different structure \cite{Partay2008,Sega2014a,bonn2015molecular} than the bulk liquid. However, a systematic analysis of the properties of water layers in the supercooled regime is missing, particularly taking care of analyzing the distribution of HDL and LDL water, which is supposed to be of particular relevance in the no-man's land.

Here, we use extensive molecular dynamics simulations of water in the supercooled regime to investigate the region around the surface tension's second inflection point with high accuracy. We characterize the local composition of the two liquid phases as a function of the proximity to the outermost molecular layer. We find that the second inflection point lies at the somewhat unexpectedly high temperature of 280 K, in correspondence with the appearance of a shoulder in the density profile and the TMD. While the presence of the shoulder could be partly explained by the reduced amplitude of thermal capillary waves, we observe the concurrent preference of HDL water to accumulate at the liquid/vapor interface. The larger freedom to perform structural reorganization at the liquid/vapor interface justifies the increased concentration of HDL water. Its larger entropy could be the origin of the increased surface tension in the no-man's land.

\section{Methods}
Our water model of choice is the TIP4P/2005 one~\cite{Abascal2005}, which reproduces the phase diagram of water with high accuracy. The TIP4P/2005 model shows the presence of the liquid-liquid phase transition~\cite{Russo2014,Abascal2010,limmer_theory_2014,Handle2017}, and two-state equations of state well explain its thermodynamic properties~\cite{Bresme2014, Singh2016}. Recently, Wang and coworkers~\cite{Wang2019} showed that the TIP4P/2005 model is compatible with the presence of a surface tension second inflection point, a feature accompanied by the appearance of a density shoulder (apophysis) in the supercooled regime.

We performed molecular dynamics simulations of water in the slab configuration using the GROMACS 2019.3~\cite{Abraham2015} simulation package.
We started from a cubic simulation box with periodic boundary conditions measuring \(5 \times 5 \times 5 \) nm\(^3\) filled with 4017 water molecules. We kept the molecular structure rigid using the SETTLE~\cite{miyamoto1992settle} algorithm and integrated the equations of motion using the leapfrog algorithm with a timestep of 1 fs. We computed the long-range contribution of dispersion and Coulomb interactions using the smooth version of the Particle Mesh Ewald algorithm~\cite{essmann1995smooth,sega_long-range_2017-2}. We employed a real-space cutoff of 1.3 nm and a relative interaction strength at the cutoff of $10^{-3}$ and $10^{-5}$ for the Lennard-Jones and Coulomb interactions, respectively, and a grid spacing of 0.15 nm. We used the Nos\'e–Hoover thermostat~\cite{Nose1984,Hoover1985} with a time constant of 2 ps for both the equilibration and production runs. 

First, we equilibrated for 10 ns 19 copies of the system at the target temperature values ranging from 198.15 to 348.15 K. After this initial phase, we extended the \(z\) box edge to 20 nm to obtain the liquid/vapor interface in slab configuration, as depicted in Fig.~\ref{Fig:HDL_render}. Subsequently, we performed an additional equilibration phase reaching up to 80 ns for the lowest temperature. During the production runs, which lasted up to 250 ns for the lowest temperature, we saved to disk for further analysis the configurations every 1 ps and energy and pressure every 0.1 ps. In total, we simulated more than 9.5 $\mu$s. The reader can find the details of equilibration and production runs in the Supplemental Information.

We analyzed the stored configurations for structural features, including profiles across the liquid slab of density, number of neighbors and HDL fraction. Our analy\-sis made extensive use of the Pytim package~\cite{Sega2018} which expands upon MDAnalysis~\cite{Michaud-Agrawal2011, Gowers2016}.
We computed the profiles of the quantities of interest across the slab using 0.2 nm wide bins. Additionally, we singled out the contributions of the first four interfacial layers\cite{sega_layer-by-layer_2015} using the Identification of Truly Interfacial Molecules (ITIM) algorithm~\cite{Partay2008} as implemented in Pytim, using a probe-sphere radius of 0.15 nm. We considered two molecules neighbors if the distance between the respective oxygen atoms is less than \(r_S = 0.35\) nm. This distance encompasses the first oxygen coordination shell reasonably well over a broad temperature range. To classify a molecule as HDL or LDL, we employed the fifth neighbor criterion (\(d_5\))~\cite{cuthbertson_mixturelike_2011} that categorizes molecules whose fifth neighbor lies farther away than \(r_S\) as LDL ones and as HDL ones otherwise.
Using the stored values of the pressure tensor, we calculated the surface tension 
\(\gamma\) using the mechanical definition
\begin{equation}\label{Eq:Mechanical_s_tens}
    \gamma = \frac{L_z}{2} \bigg[\langle P_{z} \rangle - \frac{\langle P_{x} \rangle + \langle P_{y} \rangle)}{2}\bigg], 
\end{equation}
where \(L_z = 20\) nm is the length of the box edge that is perpendicular to the macroscopic interface and \(\langle P_{\alpha} \rangle\) is the diagonal component of the pressure tensor in direction $\alpha$, averaged in the canonical ensemble (angular brackets) over the complete production run.

\section{Results and Discussion}\label{Ch:Results}
\begin{figure}[t]
    \centering
    \makebox[\columnwidth][c]{\includegraphics[width=1\columnwidth]{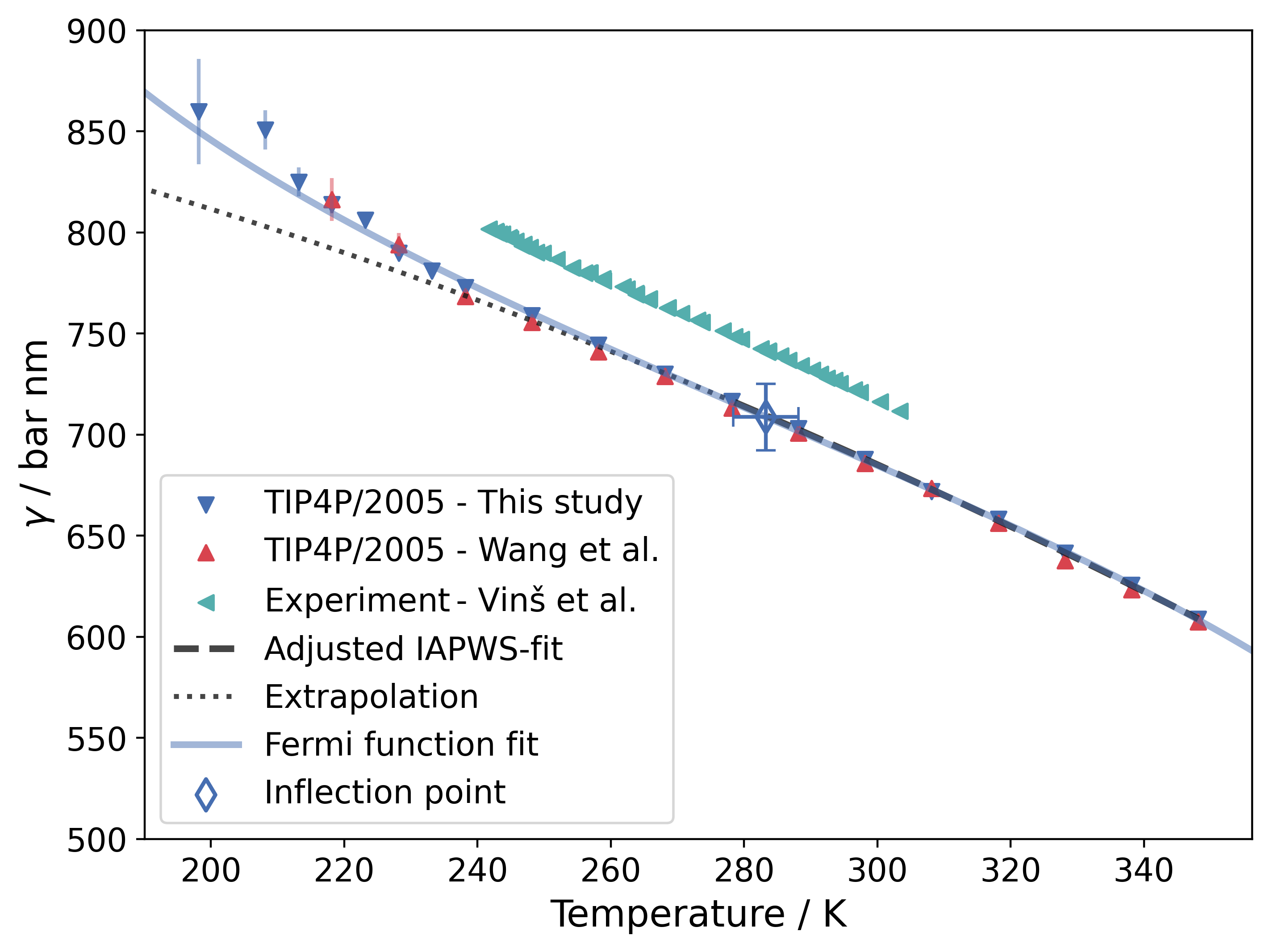}}
    \caption{Surface tension as a function of temperature for this study, Wang et al.~, Ref.~\onlinecite{Wang2019} and for Vinš et al., Ref.~\onlinecite{Vins2020}.}   \label{Fig:Plot_Surface_Tensiont}
\end{figure}

To investigate the presence of the second inflection point of \(\gamma(T)\) we considered (a) the departure of  \(\gamma(T)\) from a fit to the IAPWS equation, (b) a fit to a Fermi function, and (c) the calculation of the surface excess entropy using finite differences.
To show the departure from the IAPWS values, we first adjusted the parameters of the IAPWS equation to the TIP4P/2005 model by fitting Eq.(\ref{Eq:IAPWS}) to the measured \(\gamma(T)\) above the melting point,  keeping only the exponent \(\mu\) fixed at 11/9. The result of the optimization are \(B \)= 206.2 mN/m, \(b = -0.557\) and a critical temperature \(T_\text{c}\) of 661.543 K. The surface tension data and the IAPWS extrapolation below the melting point is shown in Fig.~\ref{Fig:Plot_Surface_Tensiont}, where we also report the simulation data of Wang and coworkers~\cite{Wang2019} and the experimental points from Ref.~\cite{Vins2020}, for comparison. The departure from the IAPWS equation shows the presence of an anomalous rise of the surface tension at low temperatures, implying the presence of an inflection point. 

To locate the inflection point accurately, we perform the best fit of our simulation data \(T(\gamma)\) to an inverted Fermi function, 
\begin{equation}\label{Eq:FermiFunction}
    T(\gamma) = \frac{C}{1 - e^{D(\gamma - \gamma_0)}} + E,
\end{equation}
with \(C,D, E\) and \(\gamma_0\) fitting parameters. This function has no particular physical significance, and it is used only as an efficient way to extract the inflection point~\cite{Sega2014a}.
Additionally, following Wang and coworkers~\cite{Wang2019}, we also calculated the surface excess entropy per unit area~\cite{rowlinson2002molecular} \(s_x = - \text{d}\gamma/\text{d} T \)
to locate the inflection point as the minimum of \(s_x\). To compute \(\text{d}\gamma/\text{d} T\) we applied the method of central differences to the sampled values of \(\gamma\).
Our data are compatible with those of Wang and coworkers~\cite{Wang2019}. Still, the larger set of temperatures and longer sampling times allow the determination of the inflection point with higher accuracy. The best fit of the Fermi function yields an inflection temperature of  283 \(\pm\) 5 K. This value is in agreement with the minimum of \(s_x\), located at 278.15 K (see SI) and is very close to the TMD (about 280 K). If we exclude the points below 223.15 K (based on possible concerns on the ergodicity of the sampled data, discussed further on), the best fit yields an inflection temperature of 275 \(\pm\) 12 K, still compatible with the estimate that uses the whole dataset.

Applying the Fermi function fit procedure to the experimental data of Vin\v{s} and coworkers leads to an estimate of the inflection point temperature of \(267 \pm  2\)~K.

Next, we investigate the possible relation of the local structure with the thermodynamics of the second inflection point in terms of density, the number of neighbors, and the fraction of HDL (or LDL) water.

\begin{figure}
    \begin{center}
    \includegraphics[width=1\columnwidth]{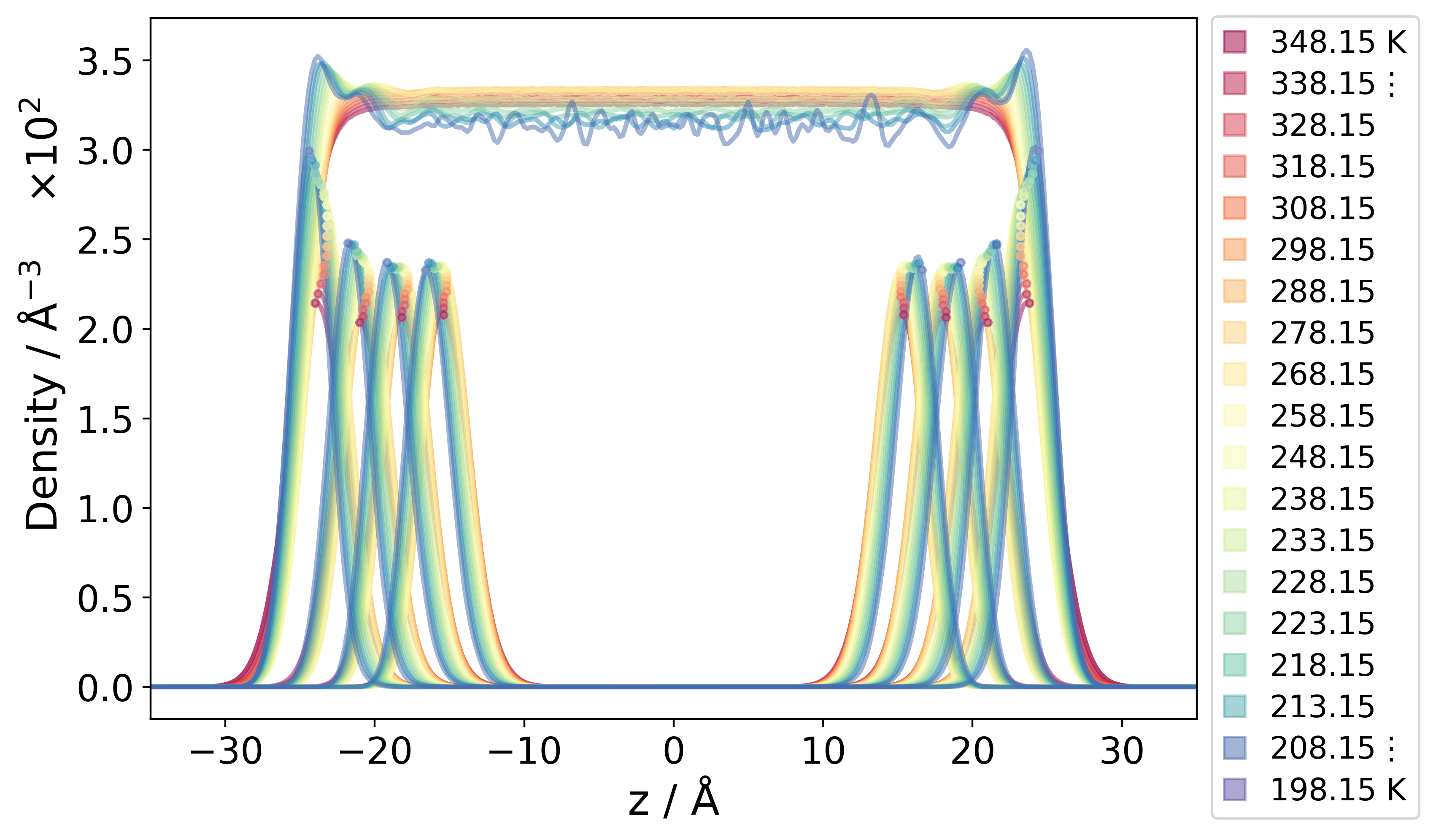}
    \end{center}
    \caption{Molecular number density profiles of the whole system and the first four layers. Dots signify layers' center of mass.
}   \label{Fig:Density_Profile}
\end{figure}

Figure \ref{Fig:Density_Profile} shows the number density profile of the whole slab and those of the first four layers on each of its two sides. A small dot marks the location of the density maximum in each layer. We notice that for the lowest 2-3 temperature values, the density profiles are not uniform far from the surface, showing signs of what seems to be insufficient sampling, despite averaging over long trajectories (250 ns for the lowest temperature). Obviously, for the lowest 2-3 temperatures, the systems show the signature of being in a glassy, non-ergodic state. However, close to the interface, the density profile appears smoother and well converged even at the lowest temperature. We will come back to this point later in our analysis. For the time being, we concentrate on the emergence of the density maximum close to the surface, which is not affected by ergodicity issues.

The density maximum at the interface is the so-called apophysis reported in several studies~\cite{Matsubara2007,Abe2014,Haji-Akbari2014,Haji-Akbari2017,malek_surface_2019,Wang2019}. 
The apophysis appears close to the maximum of the first molecular layer. Next to the apophysis, it is possible to spot a weaker but still well-defined second peak located in the proximity of the second molecular layer.
One could legitimately question whether the apophysis is only the result of the diminished amplitude of thermal capillary waves. A density profile is, in fact, a correlation function in disguise, where the local density is correlated with the location of the center of mass of the slab. Upon cooling, the smearing effect of capillary waves diminishes, and the correlation due to molecular layering can appear. Wang and coworkers~\cite{Wang2019} performed a finite size study using boxes with a transverse size $L$ of 5 and 10 nm and showed that the apophysis does not disappear when simulating the wider box. However, one needs to consider that at those low temperatures the width of the interface superimposed by thermal capillary fluctuations as predicted by the capillary wave theory~\cite{buff1965interfacial} is still smaller than the molecular size of water (0.35 nm) even for $L=10$ nm. In this sense, we could expect that capillary wave fluctuations cannot smear the apophysis even in the larger simulation box considered in Ref.~\cite{Wang2019}. 

\begin{figure}
   \includegraphics[width=.9\columnwidth]{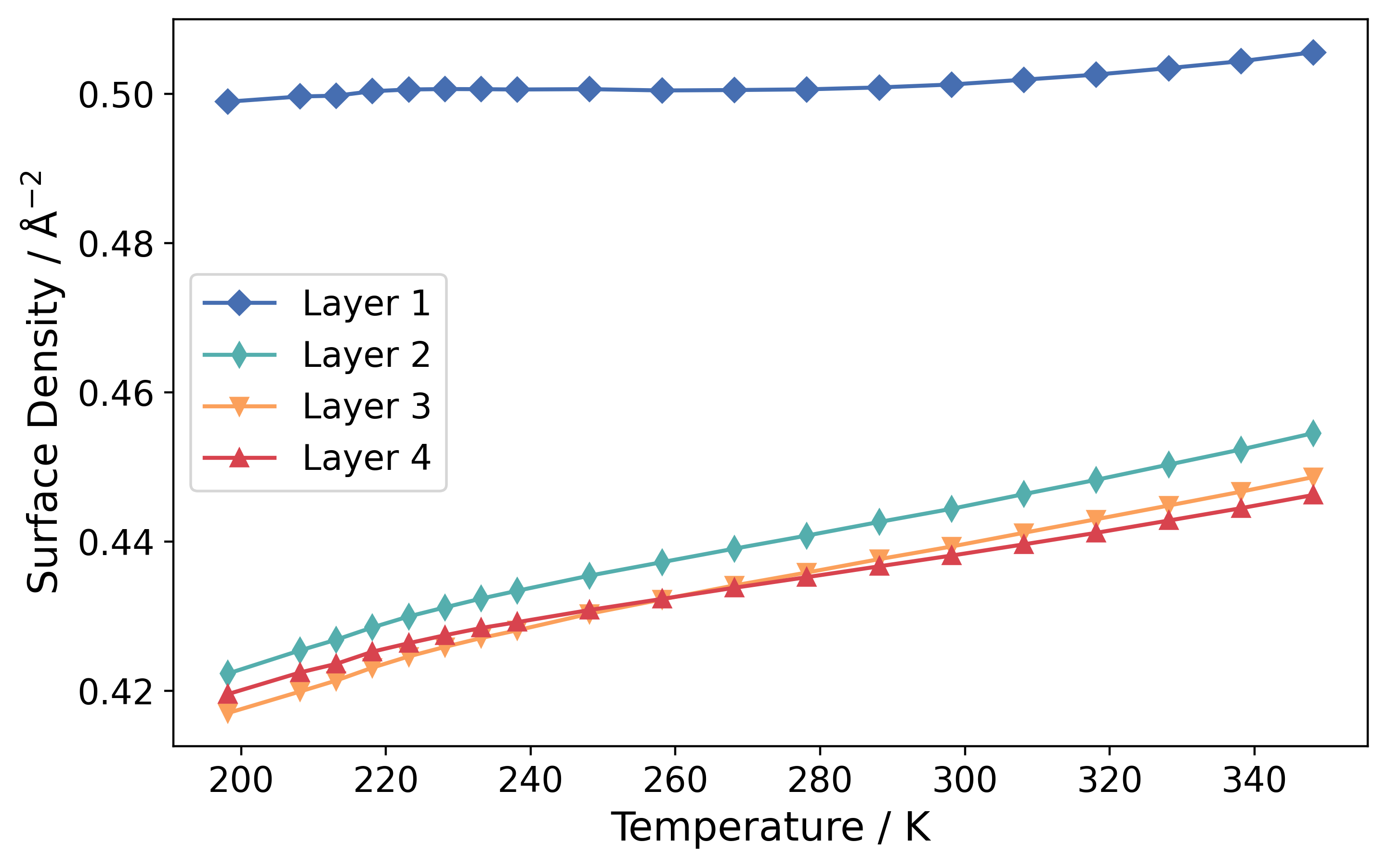}
    \caption{Integrated density profiles of the layers.}   
    \label{Fig:surface-layer-density}
\end{figure}

Despite this, the analysis of the molecular layers shows that the peak density of the first layer behaves quite differently from the subsequent ones, as its value keeps growing at a much stronger rate than the subsequent ones. When the temperature decreases, the distribution of molecules in layers changes not only in height but also in width. Integrating these distributions yields the number of molecules in each layer, which we report as surface density in Fig.~\ref{Fig:surface-layer-density}. The first layer stands out not only because of its more (laterally) densely packed structure but also because of its temperature-insensitive nature. As the layer surface density reported in Fig.~\ref{Fig:surface-layer-density} is an intrinsic property, it shows that the apophysis (or at least part of it) is not an artifact of capillary fluctuations. This confirms by other means and for planar interfaces the result obtained by Malek, Poole and Saika-Voivod for water nanodroplets~\cite{Malek2018}, who employed the Voronoi tessellation to compute the local density and avoid the spurious layering.

\begin{figure}
\begin{center}
     \includegraphics[width=\columnwidth, clip, trim=0 0 20 20]{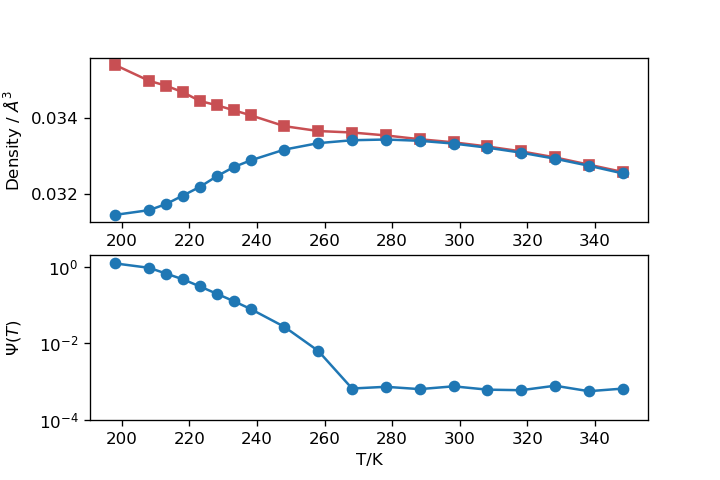}
\end{center}
    \caption{Top panel: Maximum of the density profile (red squares) and density in the middle of the slab (blue circles), as a function of temperature.
    Bottom panel: HDL peak order parameter \(\Psi\) of the HDL fraction as a function of temperature.
    \label{Fig:densitymax}}
\end{figure}

A comparison of the density value at its maximum and in bulk, as reported in Fig.~\ref{Fig:densitymax}, shows that also the apophysis starts emerging at about 280 K, roughly the same temperature as that of the inflection temperature and of the TMD. The existence of the TMD in water can be explained by the thermodynamic competition between LDL and HDL water~\cite{Poole1992,mishima2010polyamorphism,Gallo2016}, with LDL water becoming increasingly more likely than HDL when the temperature decreases. Therefore, it makes sense to investigate the water slab composition in terms of LDL and HDL fractions as a function of the proximity to the interfacial layer.

\begin{figure}
    \begin{center}
    \includegraphics[width=.9\columnwidth]{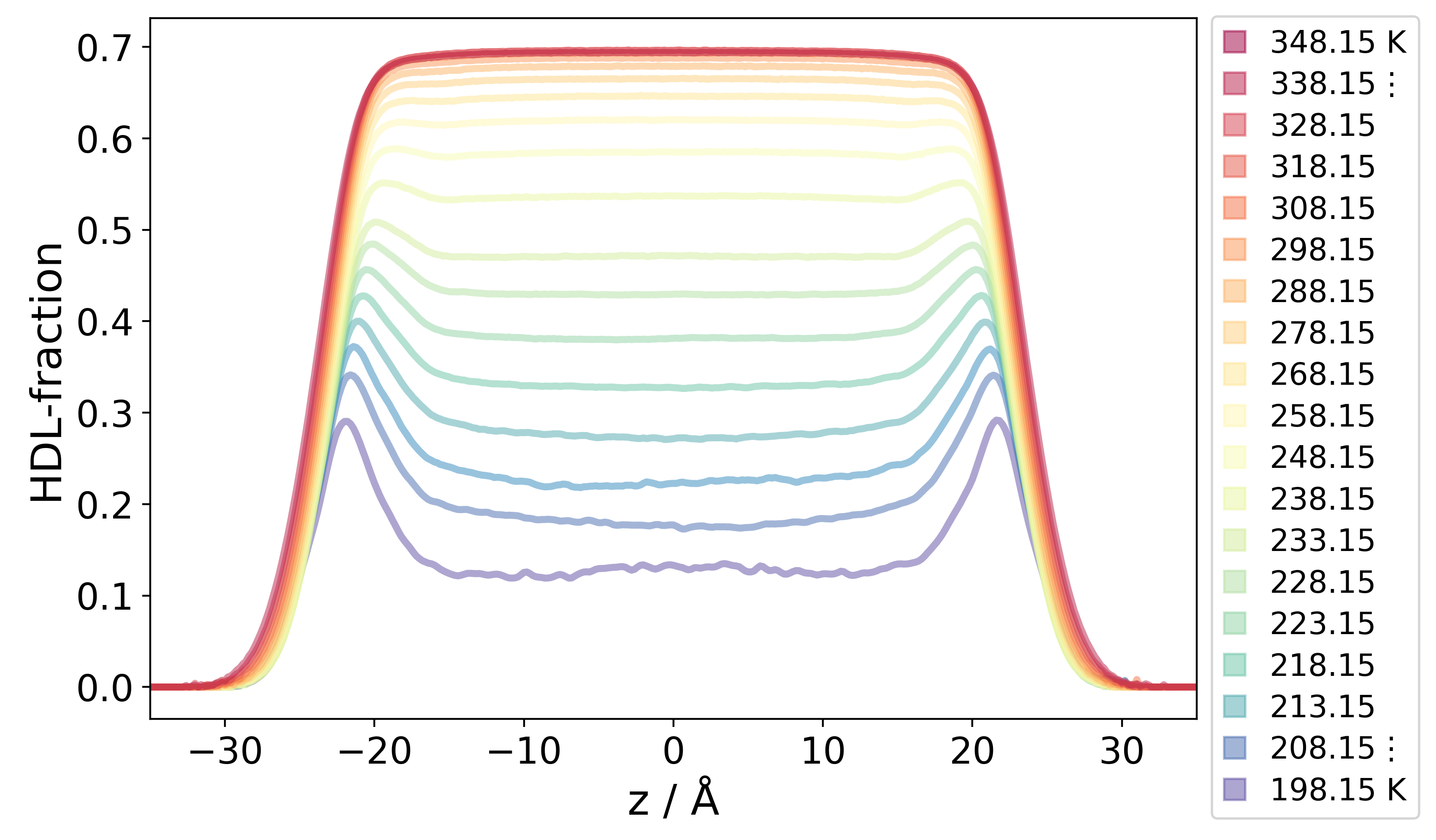}
    \end{center}
    \caption{Profiles of the fraction of HDL across the system.  }
    \label{Fig:HDL_frac}
\end{figure}

Using the fifth neighbor criterion, we compute the profile of HDL water's fraction \(f(z)\) along the surface normal. Close to the vapor phase, a growing number of molecules belonging to the first layer contribute to the histogram, and the definition of HDL fraction ceases to be meaningful. Surface layer molecules necessarily have fewer neighbors than molecules in the subsequent layers, so the HDL fraction profile drops to zero. As long as a complete shell of neighbors surrounds a molecule, starting from molecules belonging to the second layer, the value of the computed HDL fraction is meaningful. The profiles of the HDL fraction reported in Fig.~\ref{Fig:HDL_frac} show the fascinating appearance of an accumulation of HDL water at the surface upon lowering the temperature. In the surface layer, the fifth neighbor criterion cannot be used anymore, and the fraction of HDL drops to zero close to the interface. However, taking into account the fact that first layer molecules are the most densely packed (Fig.~\ref{Fig:surface-layer-density}), the emerging picture is that of a positive excess density of HDL (beyond the first layer) and HDL-like molecules (in the first layer).

To formalize the onset of the positive excess density of HDL, we compute an order parameter \(\Psi\), shown in the bottom panel of Fig. \ref{Fig:densitymax} from the HDL fraction value at its maximum \(f_\mathrm{max}\) and in the bulk \( f_\mathrm{bulk} \) as 
\begin{equation}
\Psi(T) = f_\mathrm{max}(T) / f_\mathrm{bulk}(T) -1.
\end{equation}
The order parameter is constant and close to zero at high temperature until about 270 K, when it increases upon decreasing the temperature. This behavior is similar to that of a second order phase transition, once again in the neighborhood of the TMD.

\begin{figure}
    \begin{center}
    \includegraphics[width=.9\columnwidth]{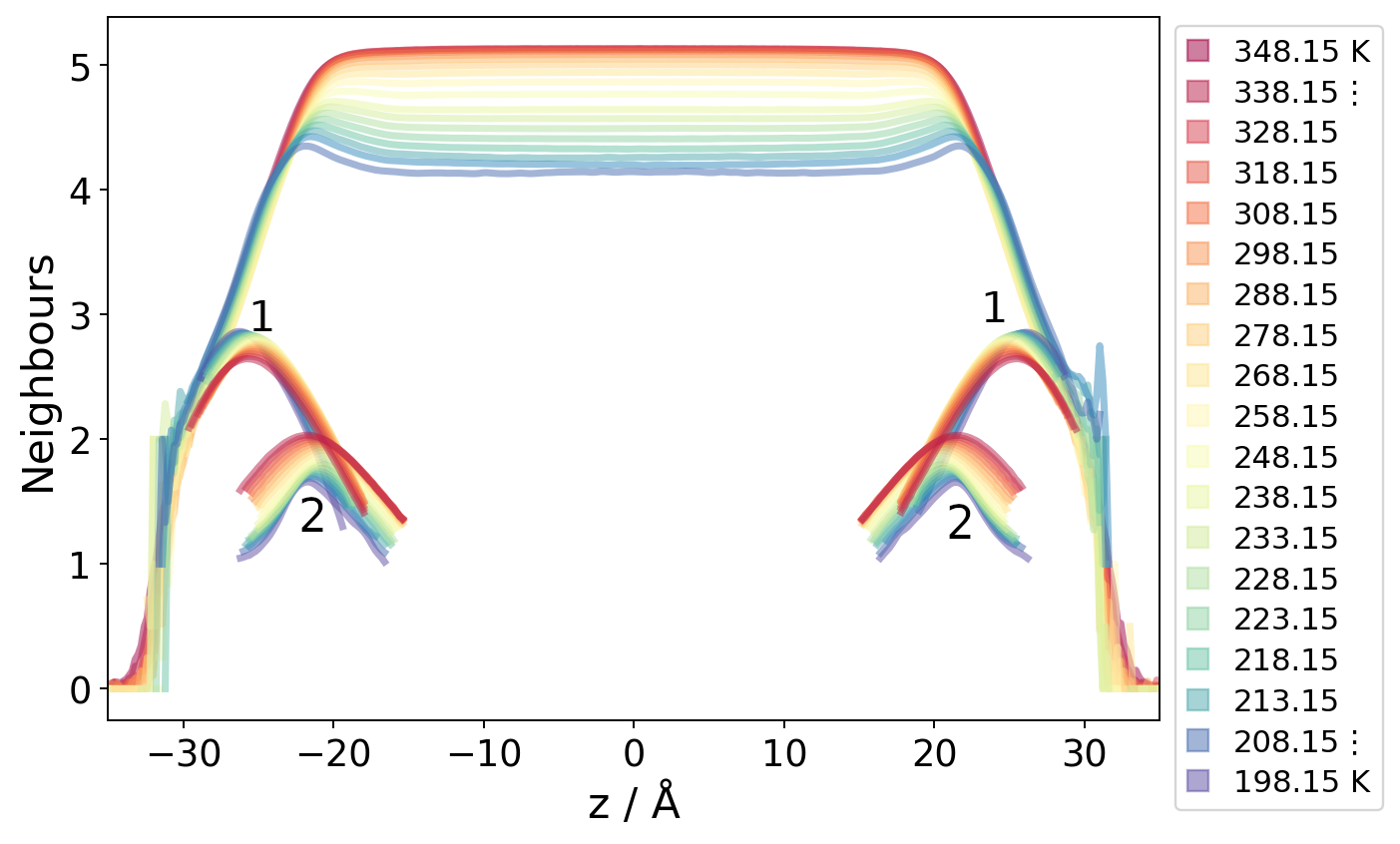}
    \end{center}
    \caption{Profiles of the number of neighbors for the whole system and of the number of intra-layer neighbors for layers 1 and 2.}
    \label{Fig:Mean_Neigh}
\end{figure}
 
It is also interesting to observe the average number of neighbors (regardless of the layer they belong to) as a function of temperature and position along the slab and the number of intra-layer neighbors for the first two layers. We report these profiles in Fig.~\ref{Fig:Mean_Neigh}. In the central region, the average number of neighbors decreases from about five at the highest temperature to slightly more than four at the lowest. This behavior underlines the change from the preponderance of HDL to that of LDL upon entering the supercooled regime. Peaks in the number of neighbors for the whole system develop at low temperatures, reflecting the presence of a positive HDL surface excess. The distributions of the intra-layer neighbors show that the second layer has about two intra-layer neighbors at high temperatures, most likely corresponding to two in-layer hydrogen bonds (the other two connecting atoms being in the first and third layers). At lower temperatures, the number of intra-layer neighbors decreases systematically. The first layer shows the opposite behavior, as the number of intra-layer neighbors increases upon cooling. This result confirms the picture of a more compact first layer shown in Fig.~\ref{Fig:surface-layer-density}. It is somehow intuitive that the possibility for molecules in the first layer to coordinate more neighbors (overall and within the same layer) is related to the larger freedom they enjoy because they do not need to fit the tetrahedral structure imposed by the hydrogen bond network in the bulk. As the system is cooled, bulk water molecules transition from HDL to LDL, lowering their configurational entropy. This transition does not happen close to the surface, where access to the less constrained first layer allows to keep a higher packing and a (relatively to the bulk) higher HDL concentration. The overall effect is an increased surface excess entropy and the consequent larger negative slope of the surface tension.

\section{Conclusions}

Using extensive molecular dynamics simulations of water along the coexistence line, we confirm and provide robust evidence of the presence of a second inflection point of its surface tension at low temperature, in connection with an increased interfacial adsorption of high density liquid water. The high accuracy of our sampling allows locating the inflection point at a higher temperature than previously suggested. The temperature of the inflection point, as estimated from the surface tension fit (\(283 \pm  5\) K)  and from its numerical differentiation (\(278\pm 5\) K) is compatible with the TMD along the coexistence line (\(278\pm5\) K). The inflection of the surface tension and its corresponding positive deviation from the fit to the IAWPS equation upon cooling correlates strongly with the appearance of a shoulder in the density profile at the liquid-vapor interface, known as the apophysis. We showed that the presence of this shoulder is not an artifact of the narrower capillary wave fluctuations at low temperatures but is indeed the result of a more compact surface layer. This compact layer is allowed to form thanks to the larger configurational entropy enjoyed by molecules in the surface layer. However, the increase in local density is not limited to the first molecular layer, as the distribution of HDL water molecules across the water slab shows. While the composition of the water slab becomes richer in LDL water, there is a clear accumulation of HDL water close to the interface when the temperature decreases. The order parameter of the HDL fraction suggests a mechanism similar to a second order phase transition, with the equivalent of a critical point located at \(268\pm 5\) K, also in proximity of the TMD.
\section*{Supplementary Material}
Additional information on the simulations and plots of surface excess entropy, number of neighbors and their profiles as a function of temperature, and peak density, HDL fraction and surface tension correlation plots.

\section*{Acknowledgments}
We acknowledge financial support of the Austrian Science Fund (FWF) through grant number I 4404. The computational results presented have been achieved using the Vienna Scientific Cluster (VSC).

%

\end{document}